\documentclass[aps,prb,twocolumn,groupedaddress,amsmath,amssymb]{revtex4}

\usepackage{graphicx}
\usepackage{dcolumn}
\usepackage{bm}
\usepackage{amsmath}
\usepackage{amssymb}
\usepackage{epsfig}
\usepackage{array}

\begin{document} \title{Optical evidence for heavy charge carriers
in FeGe} \author{V.\ Guritanu} \author{D.\ van der Marel}
\author{J.\ Teyssier} \author{T.\ Jarlborg} \affiliation{DPMC, University of Geneva, 24,
Quai E.-Ansermet, 1211 Geneva 4, Switzerland} \author{H. \ Wilhelm}
\affiliation{Diamond Light Source Ltd, Diamond House, Chilton,
Didcot, Oxfordshire, OX11 0DE, United Kingdom} \author{M. \ Schmidt,
F. \ Steglich} \affiliation{Max Planck Institute for Chemical
Physics of Solids, 01187 Dresden, Germany} \date{\today }\texttt{}

\begin{abstract}

The optical spectrum of the cubic helimagnetic metal FeGe has been
investigated in the frequency range from 0.01 - 3.1 eV for different
temperatures from 30 K to 296 K. The optical conductivity shows the
evolution of a low energy (0.22 eV) interband transition and the
development of a narrow free carrier response with a strong energy
and temperature dependence. The frequency dependent effective mass
and scattering rate derived from the optical data indicate the
formation of dressed quasi-particles with a mass renormalization
factor of 12. Similar to FeSi the spectral weight in FeGe is not
recovered over a broad frequency range, an effect usually attributed
to the influence of the on-site Coulomb interaction.

\end{abstract}

\maketitle

Cubic FeGe is a good metal at low temperature, which undergoes a
transition to helimagnetic order~\cite{wappling1968} at $T_C$ = 280
K with the magnetic moment at the iron sites of $1 \mu_{B}$. The
helix changes its orientation in a temperature interval $T_2$ $\pm$
20 K and shows pronounced temperature hysteresis~\cite{lebech1989}
between 211 K and 245 K. This material crystallizes in the B20
structure and the cubic space group P2$_1$3 lacking a center of
symmetry which is responsible for this long range order. The
iso-electronic compound FeSi has the same crystal structure. It has
a large magnetic susceptibility at room temperature, which vanishes
as the temperature approaches zero due to a small (70 meV)
semiconductor gap at E$_F$. A continuous series FeSi$_{1-x}$Ge$_{x}$
can be formed, where the metal insulator transition~\cite{yeo2003}
occurs for $x\approx0.25$. Theoretical models which have been
proposed to explain this behavior, invoke
disorder\cite{jarlborg2004}, narrow bands and different ways of
incorporating electron
correlations\cite{jaccarino1967,takahashi1979,aeppli1992,mandrus1995}.
The temperature dependent closing of the gap has been explained as a
result of a correlation gap using a two-band Hubbard
model\cite{anisimov1996,urasaki2000}, and excellent agreement was
obtained with optical
data\cite{schlesinger1993,damascelli1997,urasaki2000} but it has
been shown that vibrational disorder, if sufficiently strong, also
closes the gap\cite{jarlborg1999}.

Anisimov {\em et al.} have predicted a magnetic-field driven
semiconductor to metal transition in FeSi$_{1-x}$Ge$_{x}$, and
argue that the difference in electronic structure between FeSi and
FeGe in essence consists of a rigid relative shift of the majority
and minority spin bands for the latter material. According to this
model the optical spectra at low energies is expected to be the
superposition of a Drude peak and an interband transition across
an energy range corresponding to the forementioned relative shift
of the majority and minority bands. Experimentally relatively
little is known about the electronic structure of FeGe, for
example no optical data have been published.

Here, we report optical measurements on a cubic FeGe single crystal
at different temperatures. The real and imaginary parts of the
dielectric function were derived from the reflectivity and
ellipsometry measurements. Optical spectra of FeGe reveal the
presence of an important interband transition at 0.22 eV and an
unusual dynamics of the free carrier charge. In order to clarify the
behavior of the optical conductivity the data were compared with
local spin-density approximation (LSDA) calculations of the
electronic structure.

Cubic FeGe single crystals were grown by chemical vapor transport
method as described in detail in Ref.~\onlinecite{richardson1967}.
The single crystals were characterized by transport, magnetic and
thermodynamic measurements. It was found that a first order phase
transition to the helimagnetic state occurs at $T_C$ = 280 K.
Optical properties of FeGe were obtained using spectroscopic
ellipsometry at 0.75 - 3.1 eV and near normal incidence reflectivity
spectra were measured in the energy range from 0.01 to 0.85 eV for
different temperatures from 30 to 296 K. The complex dielectric
function $\epsilon(\omega) = \epsilon_{1}(\omega) + 4\pi
i\sigma_1(\omega)/\omega$ on a broad frequency spectral range
between 0.01 - 3.1 eV was calculated by combining the two sets of
data and using a variational Kramers-Kronig constrained
analysis~\cite{kuzmenko2005}.

\begin{figure}[bht]
\centerline{\includegraphics[width=6.5cm,clip=true]{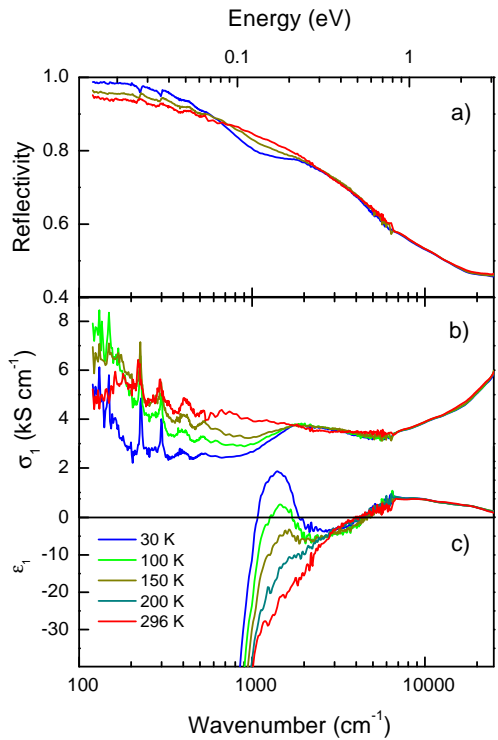}}
\caption{(Color online) The normal-incidence reflectivity
$R(\omega)$ (a), optical conductivity $\sigma_{1}(\omega)$ (b) and
real part of the dielectric function $\epsilon_{1}(\omega)$ (c) of
FeGe derived from the ellipsometry and reflectivity measurements
at several temperatures ranging from 30 to 296 K on a logarithmic
scale.}
\label{Fig_Refl_Cond_E1}
\end{figure}

Figure \ref{Fig_Refl_Cond_E1} shows the reflectivity $R(\omega)$,
the optical conductivity $\sigma_1(\omega)$, and the dielectric
function $\epsilon_1(\omega)$ of single crystalline cubic FeGe at
a serie of temperatures from 30 to 296 K over a broad frequency
range. Absolute values of the reflectivity were obtained by
calibrating the instrument using a gold-layer deposited {\em
in-situ} on the sample surface, without breaking the vacuum and
without moving the sample. This calibration procedure is designed
to fully compensate the frequency dependence of the instrument and
the geometry of the sample. Repeated experiments showed that two
weak features at 400 cm$^{-1}$ and 530 cm$^{-1}$ are not
reproducible. They coincide with the frequencies of strong
interference fringes of the thin polyethylene cryostat window. The
fact that they are not fully removed by the calibration procedure
is due to a gradual evolution of the window properties during the
time (several hours) lapsed before and after depositing the gold
layer on the crystal.

We observe that at low frequency the reflectivity increases with
decreasing temperature. In contrast, the range between 500
cm$^{-1}$ and 1500 cm$^{-1}$ $R(\omega)$ is strongly suppressed
and a dip develops at low temperature. The two distinct sharp
excitations in the far infrared spectrum (230 cm$^{-1}$, 290
cm$^{-1}$) are due to optically active
phonons~\cite{damascelli1997}. Consequently the far infrared
region of the optical conductivity spectrum displayed in Fig.
\ref{Fig_Refl_Cond_E1}b is strongly temperature dependent. At low
temperature $\sigma_{1}(\omega)$ shows a minimum around 0.1 eV,
which vanishes at T$_C$. Below T$_C$ a peak in the optical
conductivity appears at 0.22 eV which looks like an onset of
interband transitions. We observe a narrowing of the free carrier
response while the temperature is lowered to zero. However a large
finite conductivity remains below the interband transition at 0.22
eV which appears to be part of free carrier contribution. The
narrowing of the free carrier response signals a strong reduction
of the scattering rate, whereas the simultaneous appearance of an
interband transition is similar to observations in Kondo-lattices
like URu$_2$Si$_2$~\cite{bonn1988},CeAl$_3$~\cite{awasthi1993},
CeCoIn$_5$ and CeIrIn$_5$~\cite{mena2005}.

The dielectric function (see Fig. \ref{Fig_Refl_Cond_E1}c) has a
zero-crossing at 0.4 eV for all measured temperatures, which we
assign to the plasma resonance of the conduction electrons. A
second zero crossing occurs below 150 K. This low frequency
crossing is strongly temperature dependent and is shifting toward
lower energy as the temperature is decreasing. Such a line shape
of $\epsilon_{1}(\omega)$ resembles the heavy fermion systems,
where the low frequency plasmon is a characteristic feature of the
heavy quasiparticles~\cite{mena2005}.

\begin{figure}[bht]
\centerline{\includegraphics[width=6.5 cm,clip=true]{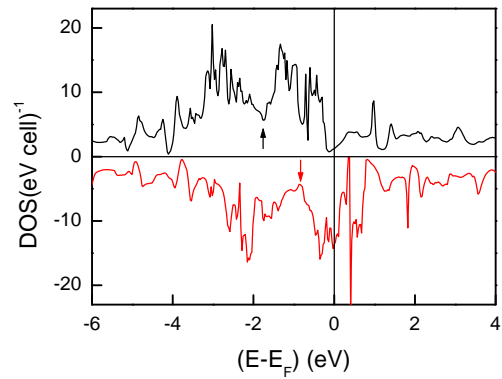}}
\caption{(Color online) Spin polarized density of states (DOS) of
FeGe. The energy is relative to $E_F$.}
\label{Fig_DOS}
\end{figure}

We calculated the electronic structure using the linear muffin-tin
orbital (LMTO) code~\cite{jarlborg2004} with the self-consistent
LSDA method, resulting in a ferromagnetic groundstate with a
magnetic moment of 1$\mu_B$ per Fe-atom. The density of states,
shown in Fig. \ref{Fig_DOS}, is consistent with the schematic
density of states of Anisimov {\em et al.} (see Fig. 3 of
Ref.~\onlinecite{anisimov2002}). The theoretical optical
conductivity has been calculated as a sum of all band transitions
within 2.7 eV of $E_F$ at 9216 k-points of the irreducible Brillouin
zone, including the dipole matrix elements. The effect of thermal
disorder and zero-point motion on the band structure is introduced
as a band broadening, which is assumed to be equal for all bands.
The result of disorder is also a smearing of the spectra at higher
energies. The parameter for the band broadening is estimated from
calculations of disordered supercell calculations for FeSi and FeGe
~\cite{jarlborg1999,pedrazzini2007}.

\begin{figure}[bht]
\centerline{\includegraphics[width=6.5
cm,clip=true]{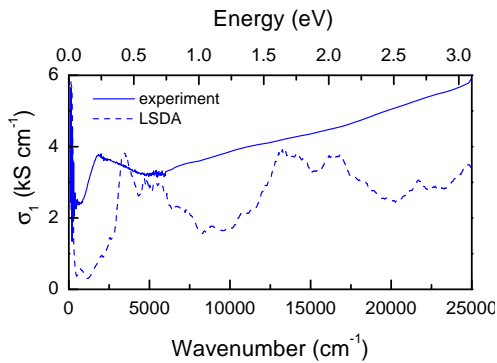}} \caption{(Color online) The optical
conductivity spectra of the helimagnetic state of FeGe.  Solid and
dashed lines denote the experimental and calculated optical
spectra, respectively.}
\label{Fig_Spectr}
\end{figure}

The comparison between the calculated and measured optical
conductivity (see Fig. \ref{Fig_Spectr}) is not as good as for
CoSi and FeSi~\cite{mena2004}. Theoretically we find the onset
weak interband transition at an energy as low as 80 meV followed
by a gradual increase of the optical conductivity to a maximum of
0.4 eV. The experimental data show a minimum at 0.1 eV followed by
a peak at 0.22 eV. In addition two peaks are predicted at 1.6 eV
and 2 eV. The experimental high frequency optical spectra is less
structured compared to the calculated one. The reason could be
that at high energy the scattering is very large. We associate the
experimental maximum at 0.22 eV with the theoretical peak at 0.4
eV and the higher energy structures to the theoretical peak
predicted at 1.6 eV. We have varied the lattice parameter in the
LSDA calculation to see if this would improve the agreement with
the experimental data, but the position of the peak at 0.4 eV
turned out to be robust.

In order to further analyze the low frequency behavior we use the
extended Drude formalism. This model is only meaningful in the
energy region where the optical response is due to mobile carriers
and not to the bound ones. The strong temperature dependence of the
optical data (Fig. \ref{Fig_Refl_Cond_E1}b) for frequencies below
0.1 eV, naturally assigned to the mobile carriers, suggest that this
model can be applied at frequencies lower than 0.1 eV. According to
this formalism the scattering rate 1/$\tau(\omega)$ and the
effective mass $m^{\ast}(\omega)/m$ are represented as follows
\begin{eqnarray}
\frac{m^{\ast}(\omega)}{m}+\frac{i}{\omega\tau(\omega)}=\frac{\omega^{2}_{p}}{\omega^{2}(\epsilon_{\infty}-\epsilon(\omega))},
\label{mstar}
\end{eqnarray}
where $\hbar\omega_{p}=4.4 eV$ is the total Drude plasma frequency
and $\epsilon_{\infty}=60$ is the high-frequency dielectric
constant, due to the bound charge polarizability. The value of
$\omega_{p}$ was obtained by integrating $\sigma_1(\omega)$ up to
the onset of interband absorption. Since there is not a complete
separation between the intraband and interband responses (see Fig.
\ref{Fig_Refl_Cond_E1}a ) the value of $\omega_p$ determined in this
way is somewhat ambiguous. However, the choice of $\omega_p$ does
not affect the frequency dependence of $\tau(\omega)$ and
$m^*(\omega)$.

\begin{figure}[bht]
\centerline{\includegraphics[width=6.5 cm,clip=true]{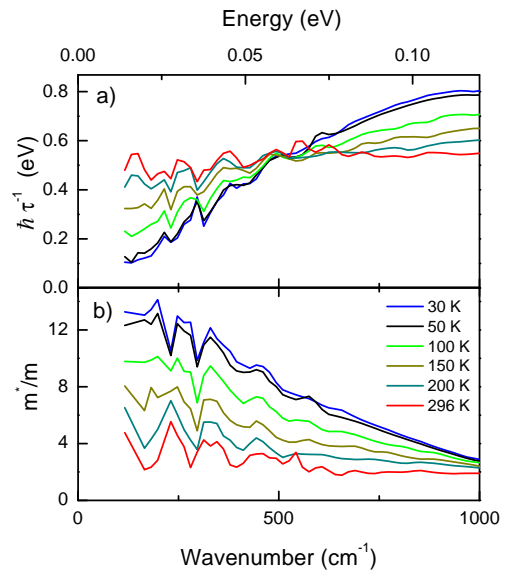}}
\caption{(Color online) Extended Drude analysis of the optical
conductivity of FeGe. The details of 1/$\tau(\omega)$ above 500
cm$^{-1}$ are due to the proximity of the oneset of interband
transitions at 0.1 eV. }
\label{Fig_Tau}
\end{figure}

Figure \ref{Fig_Tau} displays the spectra of scattering rate and
effective mass as a function of frequency for different
temperatures. Note that at room temperature both the scattering
rate and the effective mass are almost constant. As the
temperature approaches the transition temperature,
1/$\tau(\omega)$ shows strong frequency and temperature
dependence. The value of the scattering rate at high temperature
becomes quite large and it seems difficult to describe the
metallic state of this material. On the other hand, at low
temperature 1/$\tau(\omega)$ is much smaller than at room
temperature. This behavior of the scattering rate reflects the
narrowing of the zero frequency peak. This suggests that the
temperature dependence of the optical response shown in Fig.
\ref{Fig_Refl_Cond_E1}b cannot be explained in terms of the simple
Drude model. In addition we notice a strong suppression of the
scattering rate upon cooling, like in MnSi~\cite{mena2004} and in
the heavy fermion compounds URu$_2$Si$_2$~\cite{bonn1988},
CeAl$_3$~\cite{awasthi1993}, CeCoIn$_5$ and
CeIrIn$_5$~\cite{mena2005}, suggesting the development of heavy
quasiparticles at low temperature. As a result a renormalized
Drude absorption due to these heavy quasiparticles is observed in
the optical spectra. In relation to this, at low frequency the
electrons are dressed by interactions giving them a large
effective mass. For $\omega\rightarrow 0$ we observe
$m^{\ast}(\omega)/m \sim 12$. At high frequency the electrons are
no longer dressed by the interaction and the effective mass
reduces to the band mass as can be seen in Fig. \ref{Fig_Tau}b. In
contrast to the case of heavy fermions systems, FeGe has no 4f
electrons but the large mass renormalization factor appears to be
of the same order of magnitude as that observed in CeCoIn$_5$ and
CeIrIn$_5$~\cite{mena2005}. In this context we speculate that FeGe
can be considered as a 3d heavy fermion system.

The half metallic ferromagnet chromium dioxide has an optical
conductivity\cite{singley1999} very similar to that of FeGe. A
suppression of $1/\tau(\omega)$ was observed below T$_C$, which
was stronger than in FeGe (Fig \ref{Fig_Tau}a). This is a natural
consequence of the halfmetallic ferromagnetism: Spin-flip
scattering, which is the dominant scattering mechanism in
ferromagnets, is completely suppressed for frequencies smaller
than the gap separating the minority bands from the Fermi level.
Note, that in {\em disordered} Co-doped FeSi, which is also a
half-metallic ferromagnet, the opposite behaviour is observed, in
that the scattering rate {\em in}creases in the magnetically
ordered state\cite{manyala2000,manyala2004,mena2006}.

\begin{figure}[bht]
\centerline{\includegraphics[width=6.5 cm,clip=true]{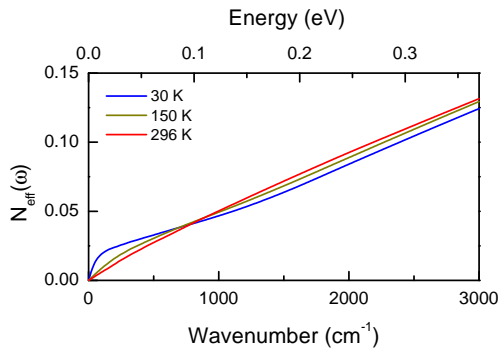}}
\caption{(Color online) The integrated spectral weight of FeGe for
a number of representative temperatures.}
\label{spectral_weight}
\end{figure}

In FeSi the spectral weight removed at frequencies below 70 meV
due to the opening of a gap at low temperatures, was observed not
to be recovered for energies up to at least 6 eV
\cite{schlesinger1993,mena2006}. Since this implies the coupling
of the conduction electrons to high energy scale excitations,
possibly on the scale of on-site Coulomb interaction, it may be an
indication that the material has features in common with a Kondo
lattice, and the insulating gap of FeSi is due strong local
electron correlation effects presumably in the 3d-shell of the
iron atoms. In view of the structural and chemical similarities
between FeSi and FeGe one might suspect electron correlation
effects to be equally important in the latter material. While FeGe
has no insulating gap at low temperatures, the optical spectra
depend strongly on temperature. In Fig. \ref{spectral_weight} the
function
%
%
$N_{eff}(\omega)= \frac{2m_eV}{\pi e^2}
\int_0^{\omega_c}\sigma_1(\omega)d\omega$
%
%
%
is displayed. For $\omega_c\rightarrow\infty$ this represents the
total number of electrons per unit of FeGe. In the region below
800 cm$^{-1}$ we observe at low temperatures that
$N_{eff}(\omega)$ increases more sharply as a function of
frequency than the high temperature data, which is a consequence
of the fact that $1/\tau$ is smaller and the Drude peak narrower
at low temperature. However, all curves cross at 800 cm$^{-1}$,
and above 1500 cm$^{-1}$ $N_{eff}(\omega,296K)$ exceeds
$N_{eff}(\omega,30K)$ by a constant amount of $\simeq 0.005$ per
FeGe formula unit. In other words: Just like in FeSi, cooling down
the sample results in a loss of spectral weight in the frequency
range 0.1 eV, which is not recovered over a broad frequency range.
While due to the smallness of the sampe the data-noise does not
permit a quantitative analysis of the spectral weight transfer for
energies larger than 0.5 eV, the temperature dependence of
$\epsilon_1(\omega,T)$ for $\hbar\omega\approx 1$ eV observed
directly with ellipsometry confirms quantitatively the trends seen
in Fig. \ref{spectral_weight}, suggesting that in FeGe, just as in
FeSi, spectral weight is redistributed over an energy range of
order 1 eV or higher when the temperature is varied. This
behaviour may have its origin in the temperature dependence of the
electron correlations resulting from the Hund's rule interaction
on the Fe-atoms\cite{schlesinger1993,urasaki2000}, in a change of
character of the bands near $E_F$ due to thermal
disorder\cite{jarlborg2004}, or in a combination of these two.

In summary, we have reported the optical properties of single
crystalline cubic FeGe, which undergoes a helimagnetic transition at
280 K. At the temperature where magnetic order occurs, a distinct
and narrow free carrier response develops, with a frequency
dependent scattering rate and a moderate mass enhancement in the
zero-frequency limit. Similar behaviour as in FeSi is observed in
FeGe, where the low energy (Drude) spectral weight appears to be
transferred to higher energies, on an energy scale of at least 1 eV,
when the material is cooled down. This, together with the observed
frequency dependent mass enhancement, indicates the important role
of electron correlations in these materials.

We gratefully acknowledge A. Kuzmenko, J. Deisenhofer and F. Carbone
for helpful discussions. This work is supported by the Swiss
National Science Foundation through grant 200020-113293 and the
National Center of Competence in Research (NCCR) 'Materials with
Novel Electronic Properties-MaNEP'.

\end{document}